\documentclass[submission,copyright,creativecommons]{eptcs}
\providecommand{\event}{SSV 2012}
\usepackage{breakurl} 
\usepackage{multirow}
\usepackage{url}

\newif \ifDraft \Draftfalse
\newif\ifFinal \Finalfalse

\ifDraft
  \newcommand{\Comment}[1]{\textbf{\textsl{#1}}}
  \newcommand{\FIXME}[1]{\textbf{\textsl{FIXME: #1}}}
\else
  \newcommand{\Comment}[1]{\relax}
  \newcommand{\FIXME}[1]{\relax}
\fi

\newcommand{\sidney}[1]{\Comment{[Sidney]#1}}

\newcommand{\question}[1]{}
\newcommand{\camera}[1]{}

\usepackage{color}

\usepackage{xspace}

\newcommand{\tool}[1]{\textsc{#1}\xspace}
\newcommand{\satabs}{\tool{SatAbs}}
\newcommand{\goanna}{\tool{Goanna}}

\usepackage{amsmath}
\usepackage{amsfonts}
\usepackage{graphicx}
\usepackage{amsthm}

\usepackage[final,formats]{listings}

\lstnewenvironment{clisting}[1]
{\lstset{
    basicstyle=\small\ttfamily,
    keywordstyle=\underbar,
    identifierstyle=,
    commentstyle=\slshape,
    stringstyle=,
    showstringspaces=false,
    keywords={EMIT, AWAIT},
    sensitive=false,
    morecomment=[l]{//},
    morecomment=[s]{/*}{*/},
    numberstyle=\tiny,
    stepnumber=1,
    numbersep=1pt,
    emphstyle=\bfseries,
    belowskip=0pt,
    aboveskip=0pt,
    #1
}}{}

\newcommand{\src}[1]{\texttt{#1}}

\newenvironment{packed_enum}{
\begin{enumerate}
  \setlength{\itemsep}{1pt}
  \setlength{\parskip}{-2pt}
  \setlength{\parsep}{0pt}
}{\end{enumerate}}

\newcommand{\ourparagraph}[1]{\vspace{1mm}\noindent\textbf{#1~~~}}

\newtheorem*{drvrule1}{Rule 1}
\newtheorem*{drvrule2}{Rule 2}
\newtheorem*{drvrule3}{Rule 3}
\newtheorem*{drvrule3-2}{Rule 3'}
\newtheorem*{drvrule4}{Rule 4}
\newtheorem*{drvrule5}{Rule 5}

\newtheorem{proposition}{Proposition}

\title{Automatic Verification of Message-Based Device Drivers}

\author{{\rm
          Sidney Amani\footnotemark[3]~\footnotemark[4] \quad
          Peter Chubb\footnotemark[3]~\footnotemark[4] \quad
          Alastair F.~Donaldson\footnotemark[5]} \\
        {\rm
          Alexander Legg\footnotemark[3] \quad
          Leonid Ryzhyk\footnotemark[3]~\footnotemark[4] \quad
          Yanjin Zhu\footnotemark[3]~\footnotemark[4]} \\
        {\footnotemark[3]~NICTA ~ ~
         \footnotemark[4]~University of New South Wales ~ ~
         \footnotemark[5]~\,Imperial College London}\\
        {sidney.amani@nicta.com.au}
        }

\def\titlerunning{Automatic Verification of Message-Based Device Drivers}
\def\authorrunning{Sidney Amani et al.}

\begin{document}

\sloppy

\maketitle

\begin{abstract}

We develop a practical solution to the problem of automatic 
verification of the interface between device drivers and the OS.  
Our solution relies on a combination of improved driver 
architecture and verification tools.  It supports drivers written 
in C and can be implemented in any existing OS, which sets it 
apart from previous proposals for verification-friendly drivers.  
Our Linux-based evaluation shows that this methodology amplifies 
the power of existing verification tools in detecting driver bugs, 
making it possible to verify properties beyond the reach of 
traditional techniques.

\end{abstract}

\section{Introduction}\label{s:intro}

Faulty device drivers are a major source of operating system (OS)
failures~\cite{Ganapathi_GP_06, Chou_YCHE_01}.  Recent studies of
Windows and Linux drivers show that over a third of driver bugs
result from the complex interface between driver and
OS~\cite{Ryzhyk_CKH_09, Ball_BCLLMORU_06}.  The OS defines 
numerous rules on the ordering and arguments of driver 
invocations, rules that often are neither well documented nor are 
stable across OS releases.  Worse, the OS can invoke driver 
functions from multiple concurrent threads, and so driver 
developers must implement complex synchronisation logic to avoid 
races and deadlocks.

In addition to causing numerous programming errors, these problems 
complicate formal analysis of device driver code.  While automatic 
verification has proved useful in detecting OS interface 
violations in device drivers,
driver verification tools remain limited in the complexity of 
properties that can be verified
efficiently~\cite{Ball_BCLLMORU_06,Cook_PR_06,Clarke_KSY_04_,Henzinger_JMNSW_02,Engler_CCH_00_}. 


One way to address the problem is through an improved device 
driver architecture that simplifies driver development and
makes them more amenable to automatic 
verification~\cite{Fahndrich_AHHHRL_06,Barnes_Ritson_09}.  In this 
architecture each driver has its own thread and communicates with 
the OS using message passing, which makes the driver control flow 
and its interactions with the OS easier to understand and analyse.
We refer to such drivers as \emph{active drivers}, in contrast to 
conventional, \emph{passive}, drivers that are structured as 
collections of entry points invoked by OS threads.

Previous implementations of active drivers in 
Singularity~\cite{Fahndrich_AHHHRL_06} and 
RMoX~\cite{Barnes_Ritson_09} OSs rely on OS and language support 
for improved verifiability.  As such, they do not help address the 
driver reliability problem in mainstream operating systems written 
in C.

In this paper we show that the benefits of active drivers can be 
achieved while writing drivers in C for a conventional OS.  To 
this end, we present an implementation of an active driver 
framework for Linux along with a new verification method that 
enables efficient, automatic checking of active driver protocols.  
Our method leverages existing verification tools for C, extended 
with several novel optimisations geared towards making active 
driver verification tractable.  Like other existing automatic 
verification techniques, the method is not complete---it helps to 
find bugs, but does not guarantee their absence.

Through experiments involving verification of several complex 
drivers for Linux, we demonstrate that our driver design and 
verification methodology amplifies the power of verification tools 
in finding driver bugs.  In particular, many properties that are 
hard or impossible to verify in conventional drivers can be easily 
checked on active drivers.

In this paper we focus on verification of active device drivers.  
A detailed account of the design and implementation of the active 
driver framework for Linux and its peformance evaluation can be 
found in the accompanying technical report~\cite{Amani_CDLRZ_12_}.

The rest of the paper is structured as follows.  
Section~\ref{s:motivation} introduces the active driver 
architecture.  Section~\ref{s:protocols} presents our visual 
formalism for specifying active driver protocols.  
Section~\ref{s:verification} describes our verification 
methodology.  Section~\ref{s:implementation} outlines the design 
and implementation of the active driver framework for Linux.   
Section~\ref{s:evaluation} presents experimental results.  
Section~\ref{s:related} surveys related work on device driver 
verification.  Section~\ref{s:conclusion} concludes the paper.

\section{Passive vs active drivers}\label{s:motivation}

In this section we discuss the shortcomings of the conventional 
driver architecture
and show how active drivers address these shortcomings.

\ourparagraph{Passive drivers} The passive driver architecture 
supported by all mainstream OSs suffers from two problems that 
complicate verification of the driver-OS interface: \emph{stack 
ripping} and \emph{concurrency}.



A passive device driver comprises a collection of entry points 
invoked by the OS.  When writing the driver, the programmer makes 
assumptions about possible orders in which its entry points are 
going to be activated; however these assumptions remain implicit 
in the implementation.
As a result, the control flow of the driver is scattered across 
multiple entry points and cannot be reconstructed from its source 
code.  This phenomenon is known as stack 
ripping~\cite{Adya_HTBD_02}.

To complicate things further, the OS can invoke driver entry 
points from multiple concurrent threads, forcing driver developers 
to implement intricate synchronisation logic to avoid races and 
deadlocks.  Multithreading further complicates automatic 
verification of device drivers, as thread interleaving leads to 
dramatic state explosion.

Previous research~\cite{Ryzhyk_CKH_09} has shown that the vast 
majority of device drivers do not get any performance benefits 
from multithreading.  The performance of most drivers is bound by 
I/O bandwidth rather than CPU speed, therefore they do not require 
true multiprocessor parallelism.  Device drivers are multithreaded 
simply by virtue of executing within the multithreaded kernel 
environment and not because they require multithreading for 
performance or functionality.

\ourparagraph{Active drivers} In contrast to passive drivers, an 
active driver runs in the context of its own thread.
Communication between the driver and the OS occurs via message 
passing.  The OS sends I/O requests and interrupt notifications to 
the driver using messages.
The driver notifies the OS about completed requests via reply 
messages.  

In an active device driver, the order in which the driver handles 
and responds to OS requests is defined explicitly in its source 
code and can be readily analysed automatically.  Since the driver 
handles I/O requests sequentially, such analysis can be performed 
without running into state explosion due to thread interleaving.


We present our instantiation of the active driver architecture for 
Linux.  Our design is based on the Dingo active driver 
framework~\cite{Ryzhyk_CKH_09}, improving upon it in two ways.  
First, Dingo's message passing primitives are implemented as C 
language extensions.  In contrast, our framework supports drivers 
in pure C.  Second, Dingo does not support automatic driver 
protocol verification.

In our framework, the driver-OS interface consists of a set of 
\emph{mailboxes}, where each mailbox is used for a particular type 
of message.  The driver exchanges messages with the OS via 
\src{EMIT} and \src{AWAIT} primitives, that operate on messages 
and mailboxes.  
The \src{EMIT} function takes a pointer to a mailbox, a message 
structure, and a list of message arguments.  It places the message
in the mailbox and returns control to the caller without blocking.
The \src{AWAIT} function takes references to one or more
mailboxes
and blocks until a message arrives in one of them.
It returns a reference to the mailbox containing the message.
A mailbox can queue multiple messages.  \src{AWAIT} always
dequeues the first message in the mailbox.  This message is 
accessible via a pointer in the returned mailbox.

\begin{figure}[t]
\begin{center}
\small

\begin{tabular}{p{0.45\linewidth}p{0.45\linewidth}}
\hline

\begin{center}
\begin{clisting}{}
mb=AWAIT(suspend,unplug,..);
if (mb==suspend) {
  dev_suspend();
  EMIT(suspend_complete,msg);
  //Bug! Uncomment to fix
  mb=AWAIT(resume/*,unplug*/);
  ...
} else if (mb==unplug) {
  ...
}
\end{clisting}
\end{center}
&

\begin{center}
\includegraphics[width=0.65\linewidth]{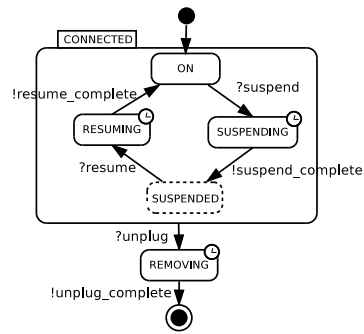}
\end{center}\\
\vspace{-9mm}\center (a) Faulty code & \vspace{-9mm}\center (b) 
Protocol
\end{tabular}
\end{center}
\vspace{-9mm}
    \caption{\label{f:active}Fragment of active driver code and
    the matching protocol specification.}
\vspace{-3mm}
\end{figure}

Figure~\ref{f:active}(a) shows a fragment of an active driver.
In line~1 the driver waits on \src{suspend} and \src{unplug} 
mailboxes.  After receiving a suspend request (checked
by the condition at line~2) the driver suspends the device 
(line~3) and notifies the OS about completion of the request by 
sending a message to the \src{suspend\_complete} mailbox (line~4).  
It then waits for a \src{resume} request at line~7.
As can be seen from this example, requests that the driver accepts 
in each state are explicitly listed in the driver source code, 
which simplifies the analysis of driver behaviour and
in particular its interaction with the OS.


\lstset{numbers=left, firstnumber=auto}

\section{Specifying driver protocols}\label{s:protocols}

This section presents our visual formalism for specifying active
driver protocols.  The formalism is similar to protocol state
machines of Dingo~\cite{Ryzhyk_CKH_09} and
Singularity~\cite{Fahndrich_AHHHRL_06}, extended with additional 
means to capture liveness and fairness constraints,
which enable the detection of additional types of driver bugs.

The active driver framework associates a protocol with each driver
interface.  The protocol specifies legal sequences of messages
exchanged by the driver and the OS.
Protocols are defined by the driver framework designer and are
generic in the sense that every driver that implements the given
interface must comply with the associated protocol.
In the case when the active driver framework is implemented within 
an existing OS,
the framework includes wrapper
components that perform the translation between the native
function-based interface and message-based active driver
protocols.


We specify driver protocols using deterministic finite state 
machines (FSMs).  The protocol state machine conceptually runs in 
parallel with the driver: whenever the driver sends or receives a 
message that belongs to the given protocol, this triggers a 
matching state transition in the protocol state machine.
Figure~\ref{f:active}(b) shows a state machine for the protocol
used by the example driver, describing the handling of power
management and hot unplug requests.  Each protocol state
transition is labelled with the name of the mailbox through which
the driver sends (`!') or receives (`?') a message.
We represent complex protocol state machines compactly using
Statecharts~\cite{Harel_87},
which organise states into a hierarchy so that several primitive states
can be clustered into a super-state.




In some protocol states the OS is waiting for the driver to
complete a request.  The driver cannot remain in such a state
indefinitely, but must eventually leave the state by sending a
response message to the OS.  Such states are called \emph{timed}
states and are labelled with the clock symbol in
Figure~\ref{f:active}(b).

In order to ensure that the driver does not deadlock in an 
\src{AWAIT} statement, the developer must rely on an additional 
assumption that if the driver waits for all incoming OS messages 
enabled in the current state, then one of them will eventually 
arrive.
This is a form of \emph{weak fairness} 
constraint~\cite{Holzmann:spin} on
the OS behaviour, which means that if some event (in this case,
arrival of a message) is continuously enabled, it will finally
occur.  Not all protocol states have the weak fairness property.  
In the protocol state machine, we show fair states with dashed 
border.  For example, the \src{SUSPENDED} state in 
Figure~\ref{f:active}b is fair, which guarantees that at least one 
of \src{resume} and \src{unplug} messages will eventually arrive 
in this state.


A protocol-compliant device driver must obey the following 5 
rules.


\begin{drvrule1}{(EMIT)} The driver is allowed to emit a        
    message to a mailbox iff this message triggers a valid state 
    transition in the protocol state machine.
\end{drvrule1}


\begin{drvrule2}{(AWAIT1)} When in a state where there is an 
    enabled incoming message, the driver must eventually issue an 
    \src{AWAIT} on the corresponding mailbox or transition into a 
    state where this message is not enabled.
\end{drvrule2}

\begin{drvrule3}{(AWAIT2)} All \src{AWAIT} operations eventually 
    terminate.  Equivalently, whenever the driver performs an 
    \src{AWAIT} operation, at least one of its protocols must be 
    in a fair state and the \src{AWAIT} must wait for all enabled 
    messages of this protocol.
\end{drvrule3}

\begin{drvrule4}{(Timed)} The driver must not remain in a timed 
    state forever.
\end{drvrule4}

\begin{drvrule5}{(Termination)} When the main driver function 
    returns, the protocol state machine must be in a final state.  
    Note that this rule does not require that every driver run 
    terminates, merely that if it does terminate then all  
    protocols must be in their final states.
\end{drvrule5}

Rules~1, 3 and 5 describe \emph{safety} properties,
whose violation can be demonstrated by a finite
execution trace.
Rules~2 and 4 are \emph{liveness} rules, for which counterexamples
are infinite runs.

Going back to the example in Figure~\ref{f:active}, we can see 
that the \src{AWAIT} statement in line~6 violates Rule~3.  This 
line corresponds to the \src{SUSPENDED} state of the protocol, 
where the driver can receive \src{unplug} and \src{resume} 
messages.  By waiting for only one of these messages, the driver 
can potentially deadlock.

\section{Verifying driver protocols}\label{s:verification}


The goal of driver protocol verification is to check whether the
driver meets all safety and liveness requirements assuming fair OS
behaviour.  We use two tools to this end:
\satabs~\cite{Clarke_KSY_04_}, geared towards safety analysis, and
\goanna~\cite{Fehnker_HJLR_06}, geared towards liveness analysis.
These tools provide complementary capabilities that, when
combined, enable full verification of many driver protocols.  
We
use \satabs to check safety rules~1, 3, and 5 and \goanna to check
liveness rules~2 and 4.  This combination works well in practice,
yielding a low overall false positive rate.  Our methodology is 
compatible with other similar tools.
We use \satabs and \goanna because our team is familiar with their 
internals and has
the expertise required to implement novel performance 
optimisations for these tools.

\subsection{Checking safety}\label{s:safety}
\satabs is an abstraction-refinement based model checker for C and
C++ for checking \emph{safety} properties. 
It is designed to perform best when checking control-flow
dominated properties with a small number of data dependencies.
Active driver protocol-compliance safety checks fall into this
category.

Given a program to verify, \satabs iteratively computes and 
verifies its finite-state abstraction with respect to a set of 
predicates over program variables.  At each iteration it either 
terminates (by discovering a bug or proving that the program is 
correct) or generates a spurious counterexample.  In the latter 
case, the counterexample is analysed by the tool to discover new 
predicates, used to construct a refined program abstraction.  
Abstraction and refinement are both fully automatic.



\satabs verifies program properties expressed as source code 
assertions.  We encode rules~1 and 3 as assertions embedded in 
modified versions of \src{AWAIT} and \src{EMIT} functions.
These functions keep track of the protocol state using a global 
state variable.  The \src{AWAIT} function simulates the receiving 
of a message by randomly selecting one of incoming mailboxes 
enabled in the current state and updating the state variable based 
on the current state and the message selected.
Similarly, the \src{EMIT} function updates the state variable 
based on the current state and the message being sent.  It 
contains an assertion that triggers an error when the driver is 
trying to send a message that is not allowed in the current state.
To verify rule~5, we append to the driver's main function a check 
to ensure that, if the driver does terminate, the protocol state 
machine is in a final state.

Our preliminary experiments show that straightforward application 
of \satabs to active drivers results in very long verification 
times.  This is in part due to the complexity of driver protocols 
being verified and in part because predicate selection heuristics 
implemented in \satabs introduce large numbers of unnecessary 
predicates, leading to overly complex abstractions.  The problem 
is not unique to \satabs.  Our preliminary experiments with SLAM~\cite{Ball_BCLLMORU_06}, 
another state-of-the-art abstraction-refinement tool, produced 
similar results.
We describe several novel strategies that exploit the properties 
of active drivers to make their safety verification feasible.  We 
believe that these techniques will also be useful in other 
software protocol verification tasks.  

\ourparagraph{Protocol decomposition}
The abstraction-refinement technique is highly sensitive to the 
size of the property being checked.  Complex properties require 
many predicates.  Since verification time grows exponentially with 
the number of predicates, it is beneficial to decompose complex 
properties into simple ones that can be verified independenly.

We decompose each driver protocol state machine into a set of much 
simpler subprotocols as a preprocessing step.  The decomposition 
is constructed in such a way that the driver satisfies safety 
constraints of the original protocol if and only if it does so for 
each protocol in the decomposition.  The following proposition 
(stated informally) gives a sufficient condition for correctness 
of decomposition.

\begin{proposition}\label{p:decomp}
Consider a protocol $P$ and its decomposition into protocols 
$P_1$, \ldots, $P_n$.  If the following conditions hold then a 
driver satisfies $P$ if and only if it satisfies each of $P_1$, 
\ldots, $P_n$:
\begin{packed_enum}
\item The regular language generated by the protocol state machine 
    of $P$ is equivalent to the intersection of languages 
    generated by $P_1$, \ldots, $P_n$.

\item There exists a bijection between fair states of $P$ and the 
    union of fair states of $P_1$, \ldots, $P_n$, such that for 
    each fair state $s$ of $P$ and the corresponding fair state 
    $s'$ of $P_i$, the set of incoming messages enabled in $s$ is 
    equal to the set of incoming messages in $s'$.
\end{packed_enum}

\end{proposition}

\begin{figure}
    \footnotesize
    \begin{tabular}{p{0.07\linewidth}p{0.11\linewidth}p{0.17\linewidth}p{0.17\linewidth}p{0.17\linewidth}p{0.17\linewidth}}
        \hline
        {\center
        \vspace{-4mm}
        \includegraphics[width=\linewidth]{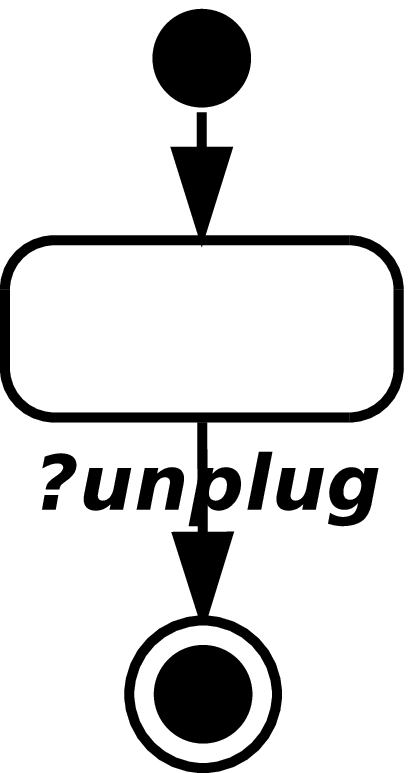}} &
        {\center
        \vspace{-4mm}
        \hspace{3mm}\includegraphics[width=\linewidth]{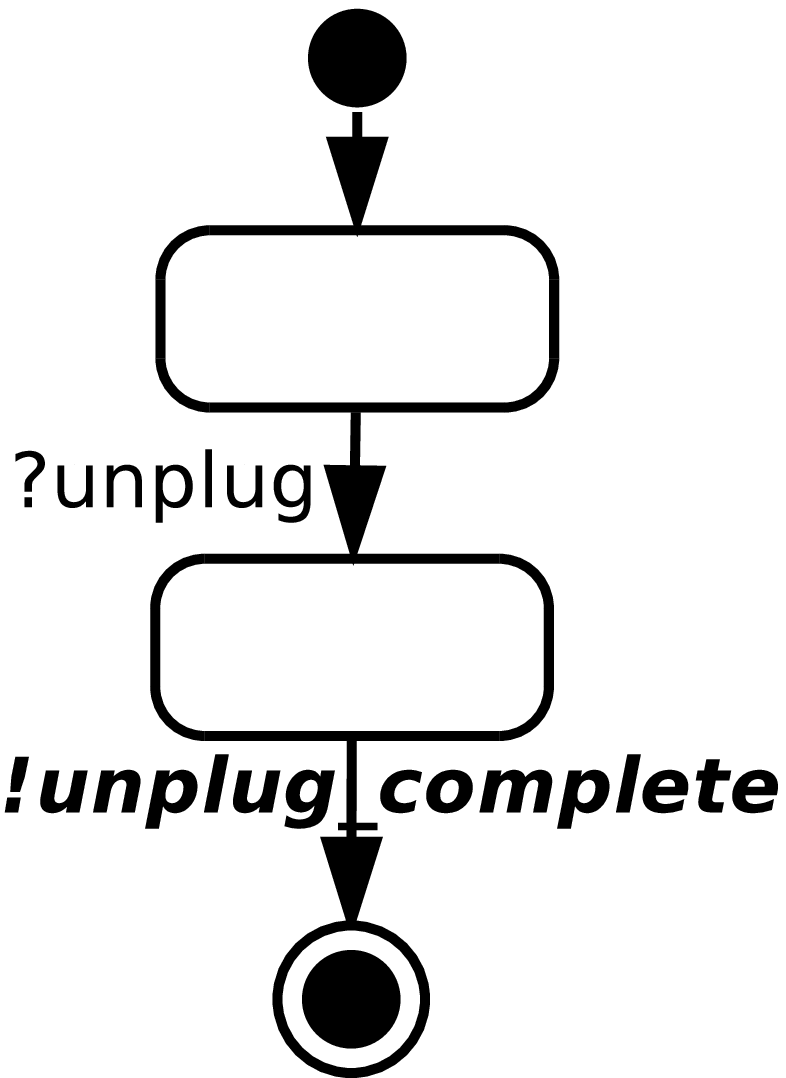}}
        &
        {\center
        \vspace{-4mm}
        \includegraphics[width=\linewidth]{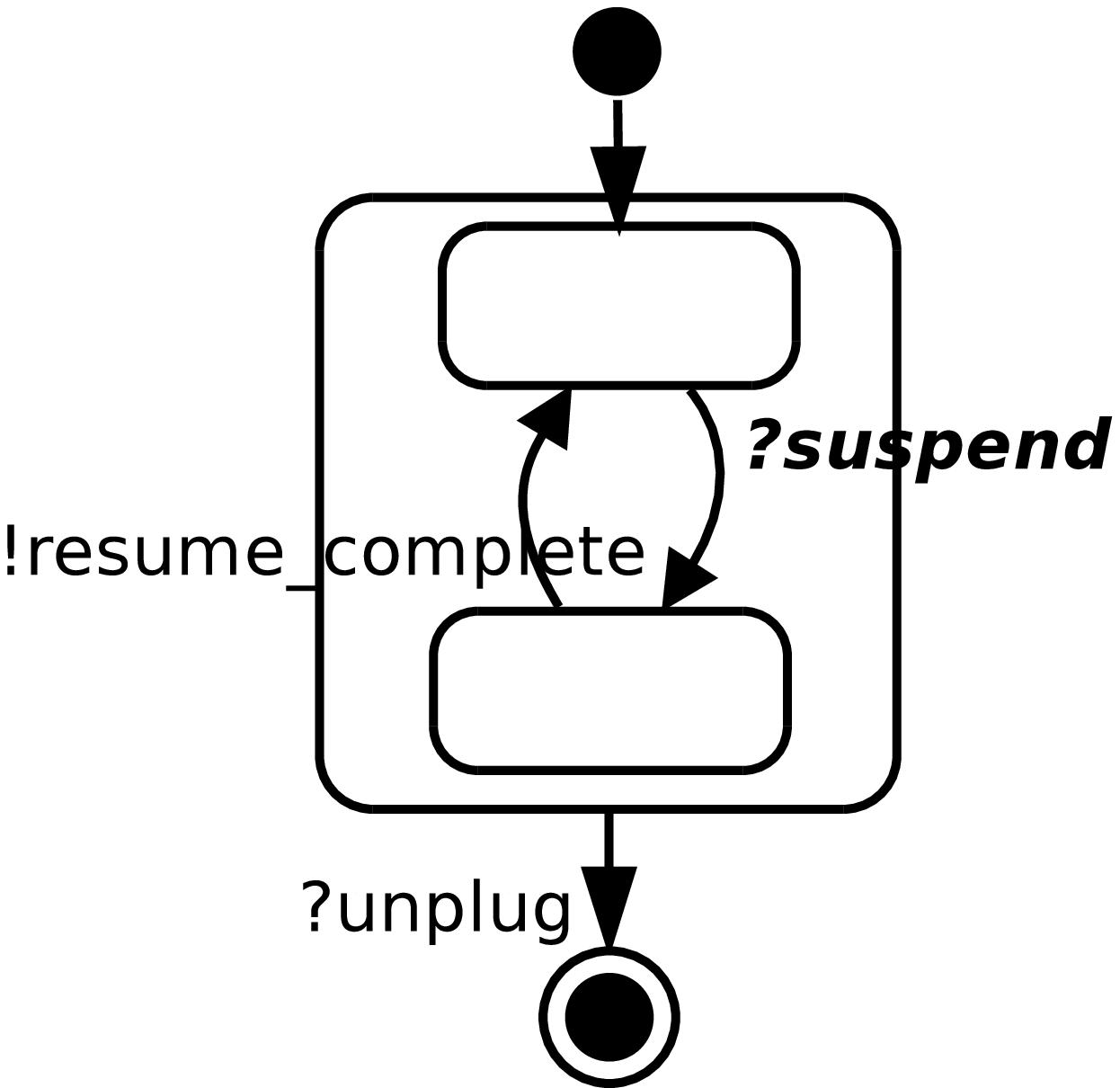}} &
        {\center
        \vspace{-4mm}
        \includegraphics[width=\linewidth]{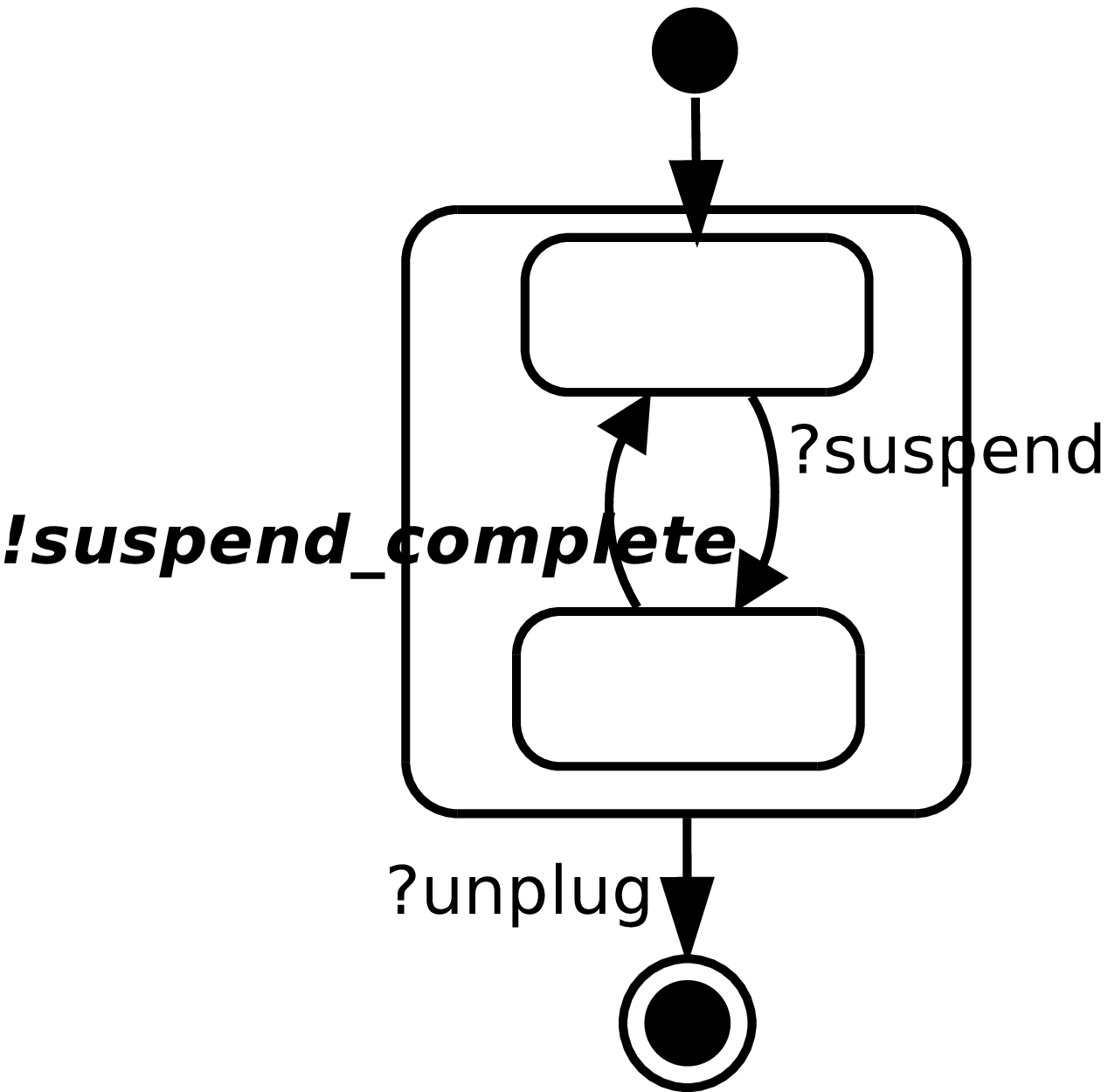}}&
        {\center
        \vspace{-4mm}
        \includegraphics[width=\linewidth]{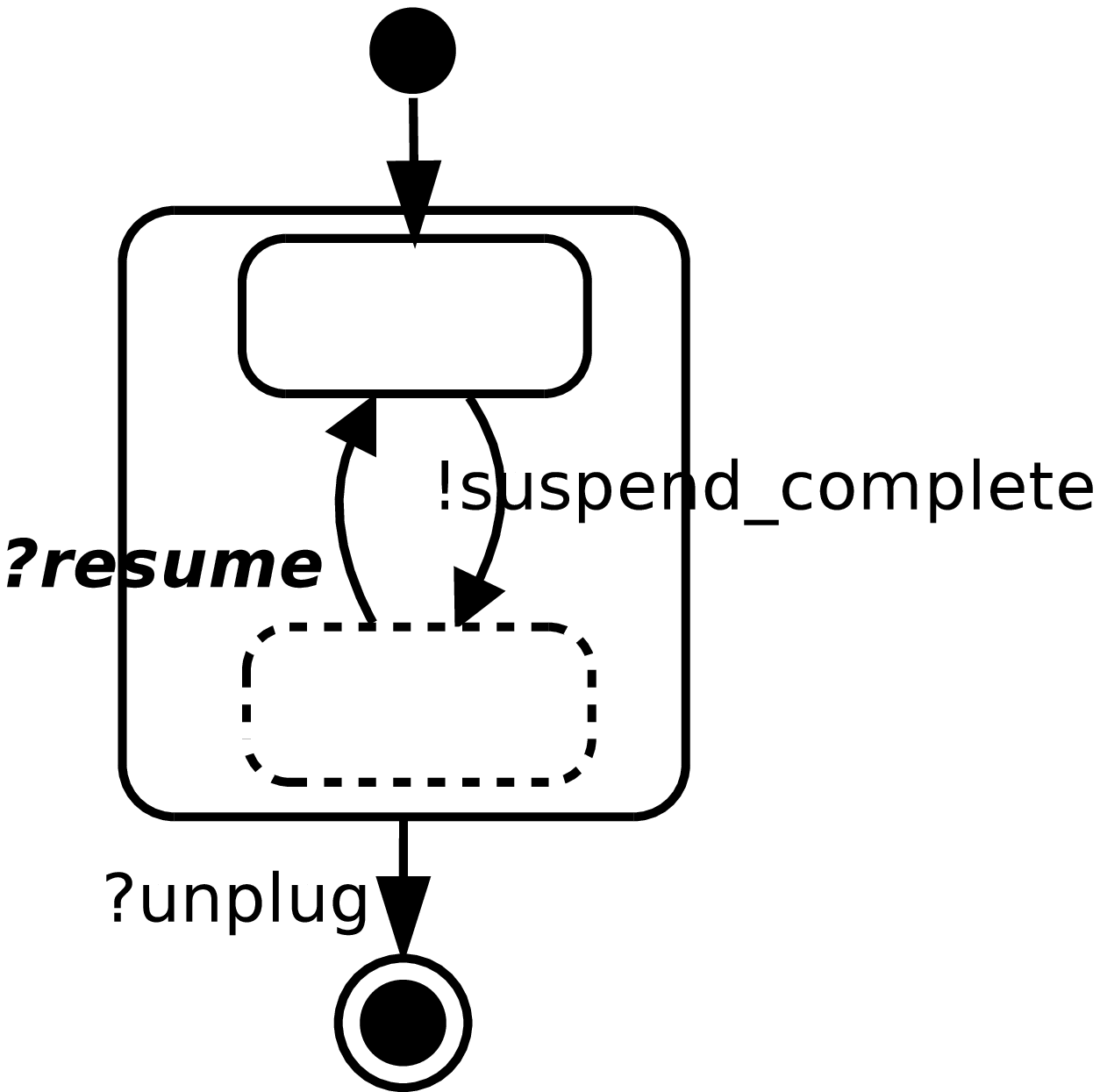}} &
        {\center
        \vspace{-4mm}
        \includegraphics[width=\linewidth]{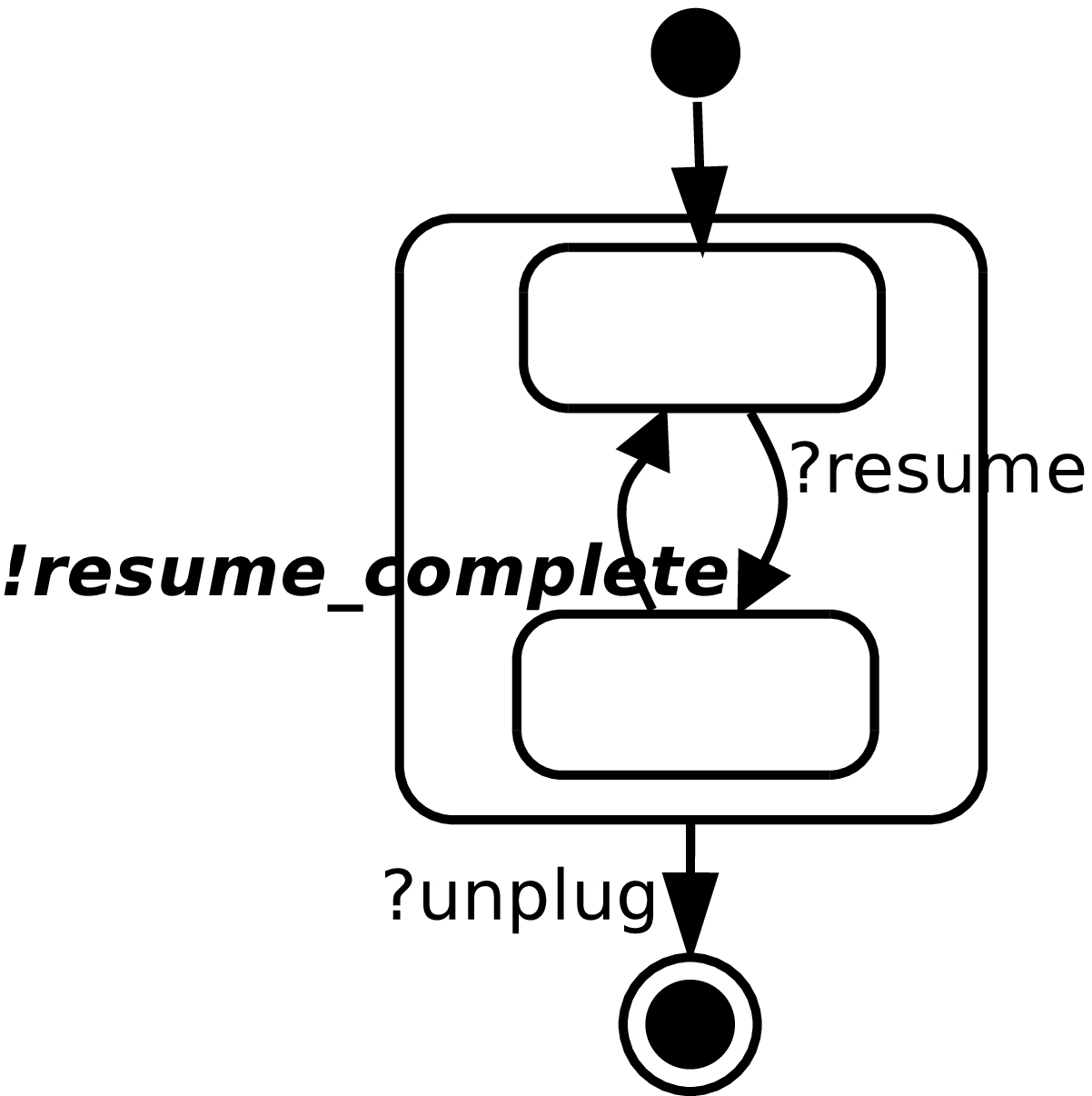}}
        \\
        \vspace{-5mm}\center\src{?unplug} &
        \vspace{-6.5mm}\center\src{!unplug\_complete}&
        \vspace{-5mm}\center\src{?suspend} &
        \vspace{-5mm}\center\src{!suspend\_complete} &
        \vspace{-5mm}\center\src{?resume} &
        \vspace{-5mm}\center\src{!resume\_complete} \\
    \end{tabular}
    \caption{\label{f:decomposition}Decomposition of the protocol
    in Figure~\ref{f:active}(b).}
\end{figure}

Figure~\ref{f:decomposition} shows one possible decomposition of 
the protocol in Figure~\ref{f:active}(b).  Every subprotocol in 
the decomposition captures a simple rule related to a single type 
of message, shown in bold italics in the diagram.  For instance,
the third protocol from the left describes the occurrence of the 
\src{suspend} message: \src{suspend} can arrive in the initial 
state, is reenabled by the \src{resume\_complete} message, and is 
permanently disabled by the \src{unplug} message.  
Messages that do not participate in the subprotocol are allowed in 
any state (as they are constrained by separate subprotocols) and 
are omitted in the diagram.

In our experience, even complex driver protocols allow 
decomposition into simple subprotocols with no more than four 
states and only a few transitions.  Verifying each subprotocol 
requires a small subset of predicates involved in checking the 
monolithic protocol, leading to exponentially faster verification.

Correctness of a decomposition can be automatically checked based 
on Proposition~\ref{p:decomp}.  Furthermore, we found construction 
of protocol decompositions to be a largely mechanical task.  As 
part of future work on the project we will investigate approaches 
to automating this task.

%
%

\ourparagraph{Automatically provide key predicates}
One way to speed-up the abstraction-refinement algorithm is to 
seed it with a small set of key predicates that avoid large 
families of spurious counterexamples.
Guessing such key predicates \emph{in general} is extremely 
difficult.  In case of active driver verification, an important 
class of key predicates can be provided to \satabs automatically.  

As mentioned above, when checking a driver protocol, we introduce 
a global variable that keeps track of protocol state.  During 
verification, \satabs eventually discovers predicates over this 
variable of the form \src{(state==1)}, \src{(state==2)}, \ldots, 
one for each state of the protocol.  These predicates are 
important to establishing the correspondence between the driver 
control flow and the protocol state machine.  We therefore provide 
these predicates to \satabs on startup, which accelerates 
verification significantly.

\ourparagraph{Control-flow transformations}
We found that it often takes \satabs many iterations to correlate 
dependent program branches.  This problem frequently occurs in 
active drivers when the driver \src{AWAIT}s on multiple mailboxes 
and then checks the returned value (e.g., line~2 in 
Figure~\ref{f:active}(a)).  If the driver executes the same 
comparison later in the execution, then both checks must produce 
the same outcome.
\satabs does not know about this correlation initially, leading to 
a spurious counterexample trace that takes inconsistent branches, 
potentially leading to spurious counteraexample traces.  These 
counterexamples can be refuted using predicate 
$p\leftrightarrow(mb==suspend)$.  In practice, however, \satabs 
may introduce many predicates that only refute a subset of these 
counterexamples before discovering $p$, which allows refuting all 
of them.



\sidney{What does the CFG transformation do in that case? just leave
        the CFG as it is? And more importantly how does the algorithm
        figure out that there are two many candidates, hardcoded maximum
        number of branches?
I Think that in general the way we solve the problem sounds too easy
we need to develop more about the implementation of the feature.
}



\begin{figure}
\center
\small
\begin{center}
    \includegraphics[width=0.65\linewidth]{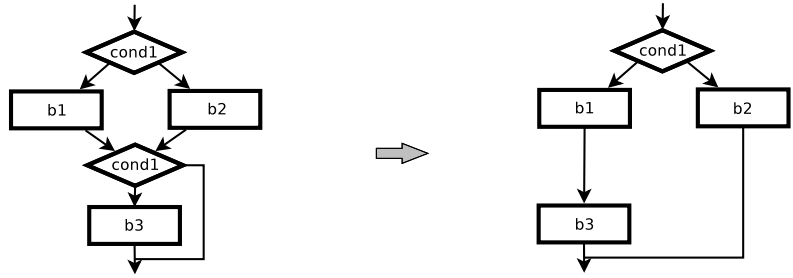}
\end{center}
\caption{\label{f:cfg}CFG transformation example.}
\end{figure}

To remedy the problem, we have implemented a novel control-flow 
graph transformation that uses static analysis to identify 
correlated branches, and merges them.  The analysis identifies, 
through inspecting the use of the \src{AWAIT} function, where to 
apply the transformation.  Then infeasible paths through each 
candidate region are identified by generating Boolean 
satisfiability queries which are discharged to a SAT solver.  The 
CFG region is then rewritten to eliminate infeasible paths.  The 
effect of the rewriting on the CFG is shown in Figure~\ref{f:cfg}.  

This technique effectively avoids the expensive search for 
additional predicates using much cheaper static program analysis.  
In our experiments, \satabs performs orders of magnitude more 
effectively over the new program structure, being able to quickly 
infer key predicates that could previously only be inferred after 
many abstraction refinement iterations and the inference of many 
redundant predicates.


\subsection{Checking liveness}\label{s:liveness}

As \satabs is restricted to analysis of safety properties, the
\goanna tool comes into play for analysis of liveness properties.
\goanna is a C and C++ bug
finding tool that supports user-defined rules written in the CTL
temporal logic~\cite{ClGrPe}, which allows natural specification
of safety and liveness properties.  
Unlike \satabs, \goanna is intended as a fast compile-time checker
and therefore does not perform data-flow analysis.

Properties to be checked for each protocol are extracted
from the protocol specification.  In particular, we apply the
\emph{AWAIT1} rule to every incoming mailbox and the \emph{Timed}
rule to every timed state of the protocol.


Describing a temporal property using the \goanna specification 
language involves two steps.  First, we identify a set of 
important program events related to the property being verified, 
such as sending and receiving of messages.  We use syntactic 
pattern matching to label program locations that correspond to 
these events.  Second, we encode the property to be checked as a 
temporal logic formula in a dialect of CTL, defined over events 
identified at the previous step.  Due to limited space, we omit 
the details of this encoding.

\section{Implementation}\label{s:implementation}

We implemented the active driver framework along with three active 
device drivers in Linux~2.6.38.  The framework consists of
loadable kernel modules and does not require any changes to other
kernel components.  The framework provides services required by 
all active drivers, including cooperative scheduling, message 
passing, and message-based interrupt delivery.  In addition it 
defines protocols for supported classes of drivers and provides 
wrappers to perform the translation between the Linux driver 
interface and message-based active driver protocols.  Wrappers 
enable conventional and active drivers to co-exist within the 
kernel.

The generic part of the framework shared by all active drivers
provides support for scheduling and message passing.  It
implements the \emph{cooperative domain} abstraction, which
constitutes a collection of cooperatively scheduled kernel threads
hosting an active driver.
Threads inside the domain communicate with the kernel via a shared
message queue.
The framework guarantees that at most one thread in the domain is
runnable at any time.  The thread keeps executing until it blocks
in the \src{AWAIT} function.  \src{AWAIT} checks whether there is
a message available in one of the mailboxes specified by the
caller and, if so, returns without blocking.  Otherwise it calls
the thread dispatcher function, which finds a thread for which a
message has arrived.  The dispatcher uses the kernel scheduler
interface to suspend the current thread and make the new thread
runnable.
In the future this design can be optimised by implementing native 
support for light-weight threads in the kernel.

\src{EMIT} and \src{AWAIT} functions do not perform memory
allocation and therefore never fail.  This simplifies driver
development, as the driver does not need to implement error
handling logic for each invocation of these ubiquitous operations.
On the other hand this means that the driver is responsible for
allocating messages sent to the OS and deallocating messages
received from the OS.
By design of driver protocols, most mailboxes can contain at most
one message, since the sender can only emit a new message to the
mailbox after receiving a completion notification for the previous
request.  Such messages can be pre-allocated statically.

Interrupt handling in active drivers is separated into top and
bottom halves.  The driver registers with the framework a top-half
function that is invoked by the kernel in the primary interrupt
context (outside the cooperative domain).  A typical top-half 
handler reads the interrupt status register, acknowledges the 
interrupt in the device, and sends an IRQ message to the driver.  
The actual interrupt handling happens inside the cooperative 
domain in the context of the driver thread that receives the IRQ 
message.  IRQ delivery latency can be minimised by queueing 
interrupt messages at the head of the message queue; alternatively 
interrupts can be queued as normal messages, which avoids 
interrupt livelock an ensures fair scheduling of interrupts with 
respect to other driver tasks.

In addition to the generic functionality described above, the
active driver framework defines protocols for supported classes of
drivers and provides wrappers to perform the translation between
the Linux driver interface and message-based active driver
protocols.  Wrappers enable conventional and active drivers to
co-exist within the kernel.

Active driver protocols are derived from the corresponding Linux
interfaces by replacing every interface function with a message or
a pair of request/response messages.  While multiple function
calls can occur concurrently, messages are serialised by the
wrapper.

Since Linux lacks a formal or informal specification of driver
interfaces, deriving protocol state machines often required
tedious inspection of the kernel source.  On the positive side, we
found that, compared to building an OS model as a C program, state 
machines provide a natural way to capture protocol constraints and 
are useful not only for automatic verification, but also as
documentation for driver developers.

Table~\ref{t:protocols} lists protocols we have specified and
implemented wrappers for.  For each protocol, it gives the number
of protocol states and transitions, and the number of subprotocols
in its decomposition (see Section~\ref{s:safety}).  
Table~\ref{t:drivers} lists active device drivers we have
implemented along with protocols that each driver supports.  All
three drivers control common types of devices found in virtually
every computer system.  These drivers were obtained by porting
native Linux drivers to the active architecture, which allows
direct comparison of their performance and verifiability against
conventional drivers.

\begin{table}
    \center
    \small
\begin{tabular}{|lccc|}
\hline
{\bf driver protocol} &
{\bf \#states} &
{\bf \#transitions} &
{\bf \#subprotocols} \\
\hline
\hline
PCI bus                         & 13 & 41 & 11 \\
Ethernet~~~~~~~                 & 17 & 36 & 6 \\
Serial ATA (SATA)               & 39 & 70 & 22 \\
Digital Audio Interface (DAI)   & 8 & 20 & 6 \\
\hline
\end{tabular}
\caption{\label{t:protocols}Implemented active driver protocols.}
\end{table}

\begin{table}
    \center
    \small
\begin{tabular}{|p{0.21\linewidth}p{0.12\linewidth}p{0.08\linewidth}p{0.08\linewidth}p{0.12\linewidth}p{0.1\linewidth}p{0.1\linewidth}|}
\hline
\vspace{-4mm}{\bf\center driver} &
{\bf supported\newline protocols} &
{\bf LOC\newline (native)} &
{\bf LOC\newline (active)} &
{\bf avg(max)\newline time(minutes)} &
{\bf avg(max)\newline refinements} &
{\bf avg(max)\newline predicates} \\
\hline
\hline
RTL8169 1Gb Ethernet             & PCI,~Ethernet      & 4,220 & 4,317 & 29 (103)  & 3 (7) & 3 (8)\\
AHCI SATA controller             & PCI,~SATA          & 2,268 & 2,487 & 123 (335) & 2 (6) & 2 (19)\\
OMAP DAI audio                   & DAI                & 583   & 705 & 5 (13)    & 2 (5) & 2 (0)\\
\hline
\end{tabular}
\caption{\label{t:drivers}Active device driver case studies,
protocols that each driver implements, size of the native
Linux and active versions of the driver in lines of code (LOC)
(measured using \src{sloccount}), along with statistics for checking safety properties
using \satabs for each driver.}
\end{table}

\section{Evaluation}\label{s:evaluation}

\subsection{Verification}

We applied the verification methodology described in
Section~\ref{s:verification} to
RTL8169, AHCI, and OMAP DAI drivers. Verification was performed on 
machines with 2GHz quad-core Intel Xeon CPUs.

\ourparagraph{Verification using \satabs and \goanna} For each of 
the three drivers we were able to verify all safety properties 
defined by their protocols using \satabs with zero false 
positives.  The last three columns of Table~\ref{t:drivers} show 
statistics for verifying safety properties using \satabs for each 
driver: average and maximum time, the number of abstraction 
refinement loop iterations and the number of predicates required 
for verification to succeed, across all subprotocols of the 
driver.  The number of predicates reflects predicates discovered 
dynamically by the abstraction refinement loop and does not 
include candidate predicates with which \satabs is initialised 
(see Section~\ref{s:safety}).


The small number of predicates involved in checking these
properties indicates that the control skeleton of an active driver
responsible for interaction with the OS has few data dependencies.  
This confirms that the active driver architecture achieves its 
goal of making the driver-OS interface amenable to efficient 
automatic verification.  At the same time, the fact that several 
refinements are required in most cases indicates that the power of 
the abstraction refinement method is necessary to avoid false 
positives when checking safety.

Despite the small number of predicates required, verification 
times are relatively high for our benchmarks.
This is due to the large size of our drivers, and the
fact that SMV~\cite{M93}, the model checker used by \satabs, was 
not designed
primarily for model checking boolean programs.  We experimented
with the BOOM model checker~\cite{DBLP:conf/tacas/BaslerHKOWZ10},
which is geared towards boolean program verification.  While in
many cases verification using BOOM was several times faster than
with SMV, we did not use it in our final experiments due to
stability issues.

All optimisations described in Section~\ref{s:safety} proved 
essential to making verification tractable.  Disabling any one of 
them led to overly large abstractions that could not be analysed 
within reasonable time.


We used \goanna to verify liveness properties of drivers as
explained in Section~\ref{s:liveness}.  \goanna performs a less
precise analysis than \satabs and is therefore much faster.
It verified all drivers in less than 1 minute while generating 8
false positives due to imprecise data flow analysis.


These results demonstrate that active drivers' protocol compliance
can be verified using existing tools.  At the same time they
suggest that an optimal combination of accuracy and verification
time requires a trade-off between full-blown predicate abstraction
of \satabs and purely syntactic analysis of \goanna.

\ourparagraph{Comparison with conventional driver verification}
In order to compare the effectiveness of our verification 
methodology against conventional verification techniques for 
passive drivers, we carried out a case study using the native 
Linux version of the RTL8169 Ethernet controller driver.  We 
analysed the history of bug fixes made to this driver,
and identified those fixes that address OS interface violation 
bugs.  A typical example involves the driver attempting to use an 
OS resource such as timer after it has been destroyed by a racing 
thread.  We found 12 such bugs.  We apply \satabs to detect these 
bugs.  \satabs has been successfully applied to Linux drivers in 
the past~\cite{Witkowski_BKW_07}.
Using \satabs as a model checker for both active and traditional 
drivers provides a fair comparison.

Detecting OS interface bugs in a passive driver requires a model 
of the OS.  We built a series of such models of increasing 
complexity so that each new model reveals additional errors but 
introduces additional execution traces and is therefore harder to 
verify.  This way we explore the best-case scenario for the 
passive driver verification methodology: using our knowledge of 
the error we tune the model for this exact error.  In practice 
more general and hence less efficient models are used in driver 
verification.

By gradually improving the OS model, we were able to find                                 
8 out of 12 bugs.  However, when being provided a model accurate 
enough to trigger the remaining 4 errors, \satabs was not able to 
find the bugs before being interrupted after 12 hours.

We carried out an equivalent case study on the active version of
the RTL8169 driver.  To this end, we simulated the 12 OS protocol
violations found in the native Linux driver in the active driver.
We were able to detect each of the 12
protocol violation bugs within 3 minutes per bug.
This result confirms that the active driver architecture along
with the verification methodology presented above lead to device
drivers that are more amenable to automatic verification than
passive drivers.

%

\subsection{Performance}

\ourparagraph{Microbenchmarks} The performance of active drivers  
depends on the overhead introduced by thread switching and message 
passing.  We measure this overhead on a machine with 2 quad-core 
1.5GHz Xeon CPUs.

In the first set of experiments, we measure the communication
throughput by sending a stream of messages from a normal kernel
thread to a thread inside a cooperative domain.  Messages are
buffered in the message queue and delivered in batches when the
cooperative domain is activated by the scheduler.  This setup
simulates streaming of network packets through an Ethernet driver.
The achieved throughput is $2\cdot{}10^6$~messages/s 
(500~ns/message) with both threads running on the same core and 
$1.2\cdot{}10^6$~messages/s (800~ns/message) with the two threads 
assigned to different cores on the same chip.  



\begin{figure}
    \center
    \includegraphics[width=0.6\linewidth]{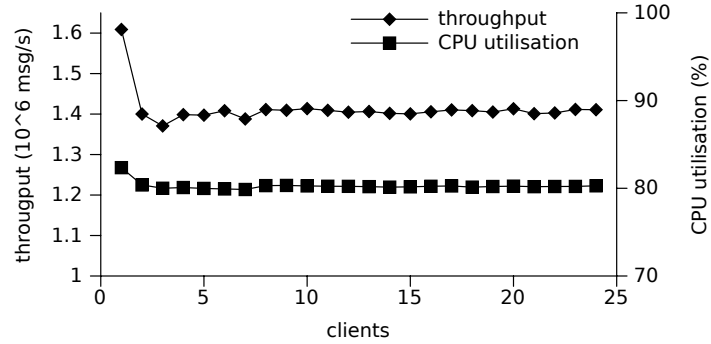}
    \caption{Message throughput and aggregate CPU utilisation over
    8 CPUs for varying number of clients.}\label{f:multiclient}
\end{figure}

Second, we run the same experiment with varying number of kernel
threads distributed across available CPU cores (without enforcing
CPU affinity), with each Linux thread communicating with the
cooperative thread through a separate mailbox.  As shown in
Figure~\ref{f:multiclient}, we do not observe any noticeable
degradation of the throughput or CPU utilisation as the number of
clients contending to communicate with the single server thread
increases (the drop between one and two client threads is due to
the higher cost of inter-CPU communication).  This shows that our
implementation of message queueing scales well with the number of
clients.

Third, we measure the communication latency between a Linux thread
and an active driver thread running on the same CPU by bouncing a
message between them in a ping-pong fashion.  The average measured
roundtrip latency is 1.8 $\mu$s.  For comparison, the
roundtrip latency of a Gigabit network link is at least
55$\mu$s~\cite{Hughes_CD_05}.

\begin{figure}[t]
\center
\footnotesize
\begin{tabular}{p{0.3\linewidth}p{0.3\linewidth}p{0.3\linewidth}}
    \includegraphics[width=\linewidth]{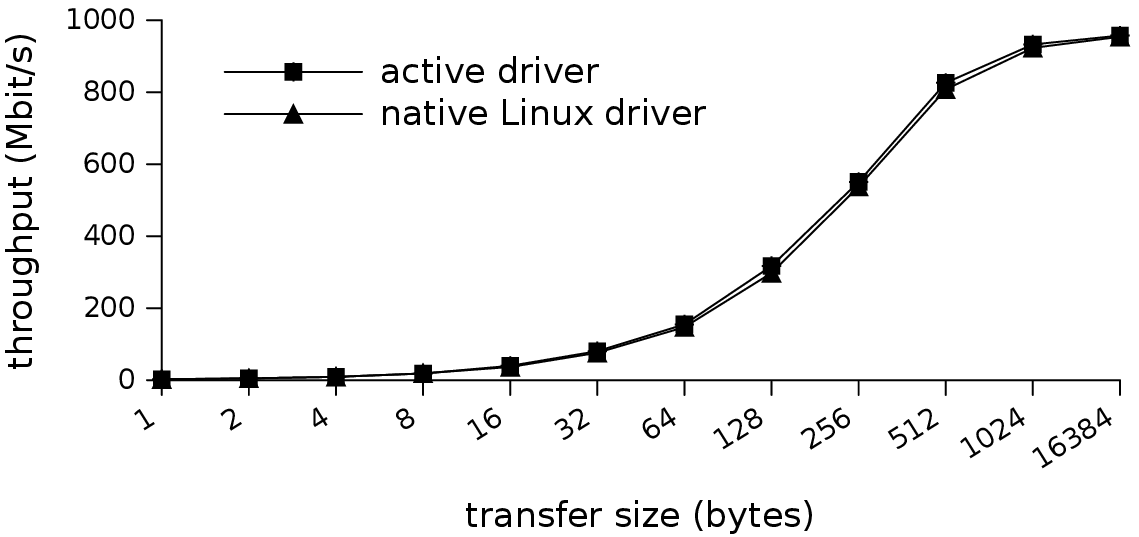}&
    \includegraphics[width=\linewidth]{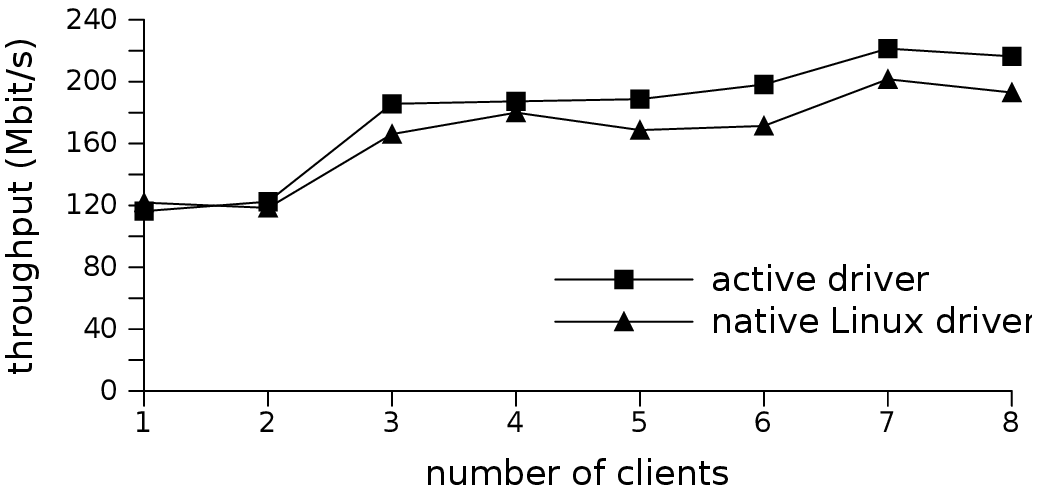}&
    \includegraphics[width=\linewidth]{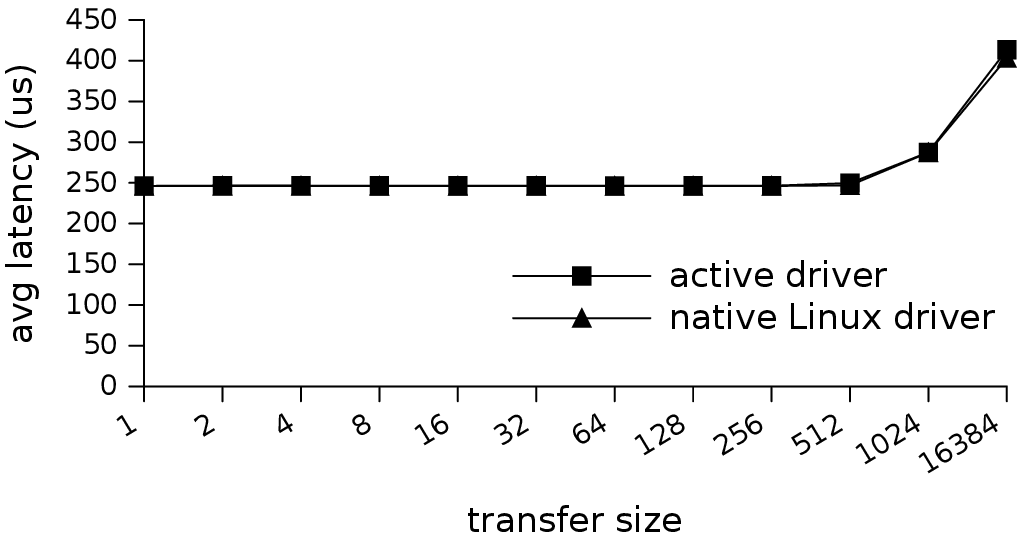}\\
    \includegraphics[width=\linewidth]{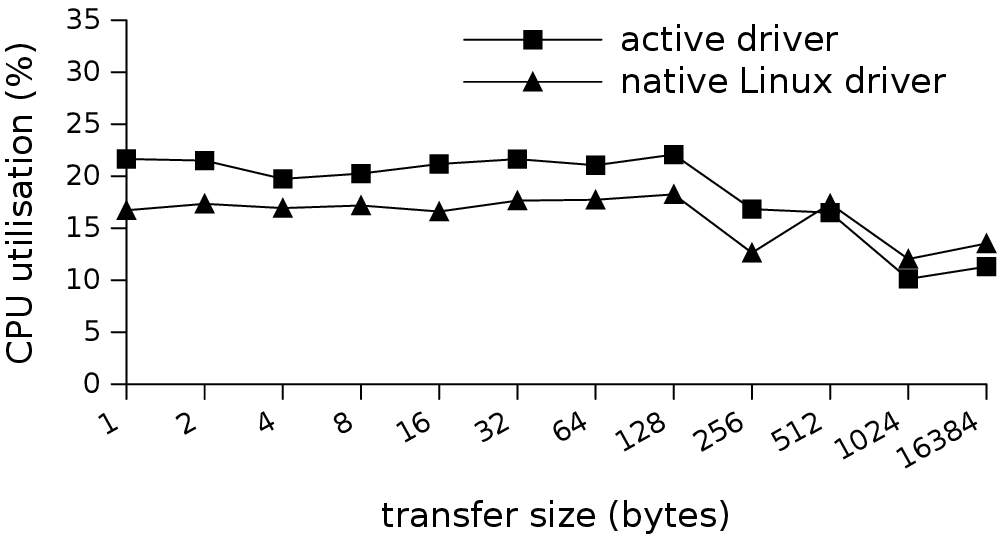}&
    \includegraphics[width=\linewidth]{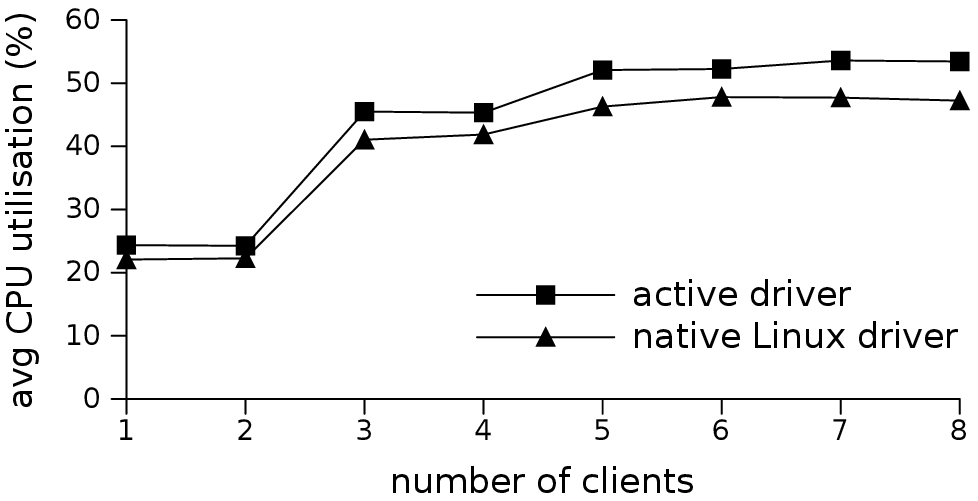}&
    \includegraphics[width=\linewidth]{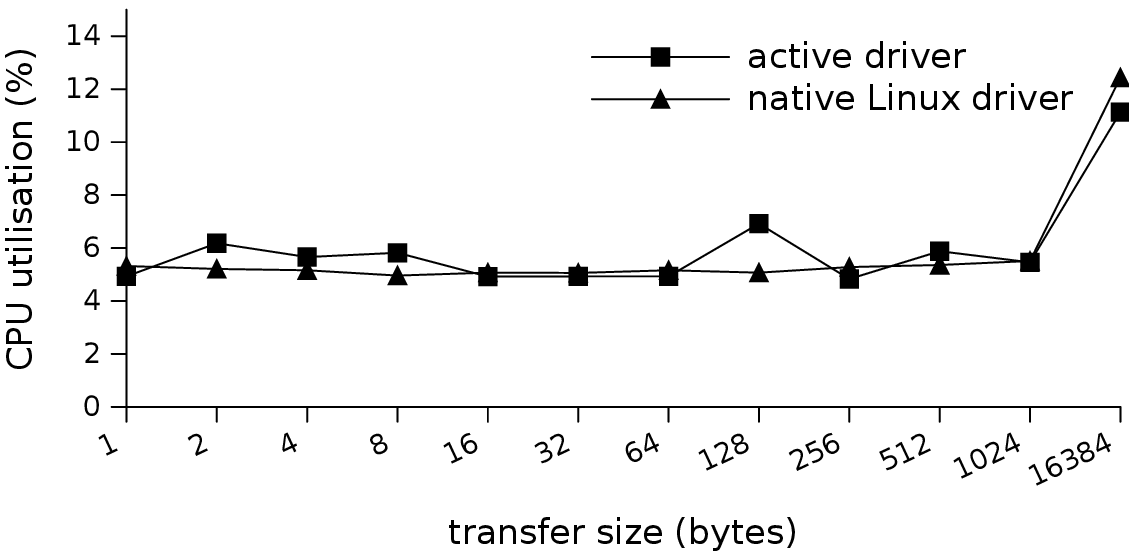}\\
    (a) UDP throughput for varying packet sizes for a single client.
        The top graph shows achieved throughput; the bottom graph shows CPU utilisation.&
    (b) UDP throughput for multiple clients (packet size=64~bytes).
        The top graph shows aggregate throughput; the bottom graph shows average CPU utilisation across 8 cores.&
    (c) UDP latency for varying packet sizes for a single client.
        The top graph shows average round-trip latency; the bottom 
        graph show CPU utilisation.\\
\end{tabular}
    \caption{\label{f:rtl}Performance of the RTL8169 Ethernet 
    driver measured with Netperf.}
\end{figure}

\ourparagraph{Macrobenchmarks} We compare the performance of the
active RTL8169 Ethernet controller driver against equivalent
native Linux driver using the Netperf benchmark suite on a 2.9GHz
quad-core Intel Core i7 machine.  Results of the comparison are
shown in Figure~\ref{f:rtl}.  In the first set of experiments we
send a stream of UDP packets from the client to the host machine,
measuring achieved throughput (using Netperf) and CPU utilisation
(using \src{oprofile}) for different payload sizes.  The client
machine is equipped with a 2GHz AMD Opteron CPU and a Broadcom
NetXtreme BCM5704 NIC.  The active driver achieved the same
throughput as the native Linux driver on all packet sizes, while
using 20\% more CPU in the worst case (Figure~\ref{f:rtl}(a)).

In the second set of experiments, we fix payload size to 64~bytes
and vary the number of clients generating UDP traffic to the host
between 1 and 8.  The clients are distributed across four 2GHz
Intel Celeron machines with an Intel PRO/1000 MT NIC.  The results
(Figure~\ref{f:rtl}(b)) show that the active driver sustains up to
10\% higher throughput while using proportionally more CPU.
Further analysis revealed that the throughput improvement is due
to slightly higher IRQ latency, which allows the driver to handle
more packets per interrupt, leading to lower packet loss rate.

The third set of experiments measures the round trip communication
latency between the host and a remote client with 2GHz AMD Opteron
and NetXtreme BCM5704 NIC.  Figure~\ref{f:rtl}(c) shows that the
latency introduced by message passing is completely masked by the
network latency in these experiments.

We evaluate the performance of the AHCI SATA controller driver  
using the \src{iozone} benchmark suite running on a system with a  
2.33GHz Intel Core~2 Duo CPU, Marvell 88SE9123 PCIe~2.0 SATA 
controller, and WD Caviar SATA-II 7200 RPM hard disk.  We run the 
benchmark with working set of 500MB on top of the raw disk.

\begin{figure}
    \center
    \includegraphics[width=0.5\linewidth]{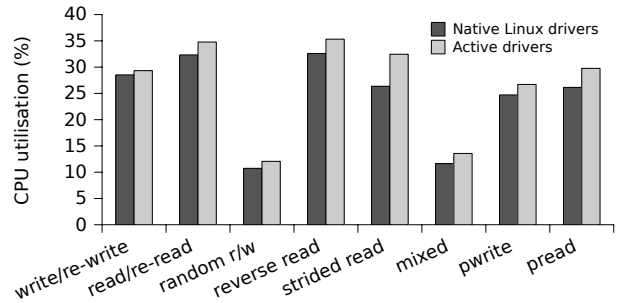}
    \caption{Native vs.\ active AHCI and ATA framework driver
    performance on the iozone benchmark.}\label{f:ahci}
\end{figure}

We benchmark the driver against
equivalent Linux driver.  Both drivers achieved the same I/O
throughput on all tests, while the active driver's CPU utilisation
was slightly higher (Figure~\ref{f:ahci}).  This overhead can be
reduced through improved protocol design.  Our SATA driver
protocol, based on the equivalent Linux interface
requires 10 messages for each I/O operation.  A clean-slate
redesign of this protocol would involve much fewer messages.

We did not benchmark the DAI driver, as it has trivial performance
requirements and uses less than 5\% of CPU.

\section{Related work}\label{s:related}

\ourparagraph{Active drivers} 
Singularity~\cite{Fahndrich_AHHHRL_06} is a research OS written
in the Sing\# programming language.  It comprises a collection of
processes communicating over message channels.  Sing\# supports a
state-machine-based notation for specifying communication
protocols between various OS components, including device drivers.
The Sing\# compiler checks protocol compliance at compile time.
RMoX~\cite{Barnes_Ritson_09} is a process-based OS written in
occam-pi.  RMoX processes communicate using synchronous
rendezvous.  Communication protocols are formalised using the CSP
process algebra and verified using the FDR tool.


The Dingo~\cite{Ryzhyk_CKH_09} active driver framework for Linux 
aims to simplify driver programming in order to help driver 
developers avoid errors.  It relies on a C language extension to 
provide language-level support for messages and threads.  Dingo 
uses a Statechart-based language to specify driver protocols; 
however it only supports runtime protocol checking and does not 
implement any form of static verification.

The \tool{CLARITY}~\cite{Chandrasekaran_CJR_07} programming 
language is designed to make passive drivers more amenable to 
automatic verification.  To this end it provides constructs that 
allow writing event-based code in a sequential style, which 
reduces stack ripping.  It simplifies reasoning about concurrency 
by encapsulating thread synchronisations inside \emph{coord} 
objects that expose well-defined sequential protocols to the user.



\ourparagraph{Verification tools}
Automatic verification tools for
C~\cite{Ball_BCLLMORU_06,Cook_PR_06,Clarke_KSY_04_,Henzinger_JMNSW_02}
is an active area of research, which is complementary to our work
on making drivers amenable to formal analysis using such tools.
Several verification tools, including SPIN~\cite{Holzmann:spin}, 
focus on checking message-based protocols in distributed systems.  
These tools work on an abstract model of the system that is either 
written by the user or extracted from the program source 
code~\cite{Holzmann_00}.
Such a model constitutes a fixed abstraction of the system that 
cannot be automatically refined if it proves too coarse to verify 
the property in question.  Our experiments show that abstraction 
refinement is essential to avoiding false positives in active 
driver verification; therefore we do not expect these tools to 
perform well on active driver verification tasks.

\section{Conclusion}\label{s:conclusion}

Improvements in automatic device driver verification cannot rely 
solely on smarter verification tools and require an improved 
driver architecture.  Previous proposals for verification-friendly 
drivers were based on specialised language and OS support and were 
not compatible with existing systems.  Based on ideas from this 
earlier research, we developed a driver architecture and 
verification methodology that can be implemented in any existing 
OS.  Our experiments confirm that this methodology enables more 
thorough verification of the driver-OS interface than what is 
possible for conventional drivers.

\section{Acknowledgements}

We would like to thank Michael Tautschnig for his help in 
troubleshooting \satabs issues.  We thank the \goanna team, in 
particular Mark Bradley and Ansgar Fehnker, for explaining \goanna 
internals and providing us with numerous ideas and examples of 
verifying active driver properties using \goanna.  We thank Toby 
Murray for his feedback on a draft of the paper.

NICTA is funded by the Australian Government as represented by the 
Department of Broadband, Communications and the Digital Economy         
and the Australian Research Council through the ICT Centre of 
Excellence program.

\end{document}